\documentclass[12pt,preprint]{aastex}

\newcommand{\simle}{\mbox{$\stackrel{<}{_{\sim}}$}}

\newcommand\vis{${\mathcal{V}^2}$}

\shorttitle{Imaging with IOTA3}
\shortauthors{Monnier et al.}
                                       
\begin{document}


\title{First results with the IOTA3 imaging interferometer:
The spectroscopic binaries $\lambda$~Vir and WR~140
}


\author{J.~D.~Monnier\altaffilmark{1}, W.~A.~Traub\altaffilmark{2}, 
  F.~P.~Schloerb\altaffilmark{3},
  R.~Millan-Gabet\altaffilmark{4}, J.-P.~Berger\altaffilmark{5},
  E.~Pedretti\altaffilmark{2},
  N.~P.~Carleton\altaffilmark{2}, S.~Kraus\altaffilmark{3}, 
  M.~G.~Lacasse\altaffilmark{2},  M.~Brewer\altaffilmark{3},  S.~Ragland\altaffilmark{2},
A.~Ahearn\altaffilmark{2},
  C.~Coldwell\altaffilmark{2}, 
P.~Haguenauer\altaffilmark{6}, P.~Kern\altaffilmark{5},  
P.~Labeye\altaffilmark{7},  L.~Lagny\altaffilmark{5}, 
  F.~Malbet\altaffilmark{5}, D.~Malin\altaffilmark{3},  
P.~Maymounkov\altaffilmark{2}, S.~Morel\altaffilmark{8},
C.~Papaliolios\altaffilmark{2}, K.~Perraut\altaffilmark{5}, M.~Pearlman\altaffilmark{2}, 
  I.~L.~Porro\altaffilmark{9},
  I.~Schanen\altaffilmark{10}, K.~Souccar\altaffilmark{3}, 
  G.~Torres\altaffilmark{2}, and G.~Wallace\altaffilmark{3} 
}
\altaffiltext{1}{monnier@umich.edu: University of Michigan (Astronomy), 500 Church St, Ann Arbor, MI 48109-1090}
\altaffiltext{2}{Harvard-Smithsonian Center for Astrophysics, 60 Garden St,
Cambridge, MA, 02138, USA}
\altaffiltext{3}{University of Massachusetts, Amherst}
\altaffiltext{4}{Michelson Science Center, California Institute of Technology, Pasadena, CA}
\altaffiltext{5}{Laboratoire d'Astrophysique de Grenoble, 414 Rue de la Piscine 38400 Saint Martin d'Heres, France}
\altaffiltext{6}{Alcatel Space Industries, Cannes, France}
\altaffiltext{7}{LETI-CEA, Grenoble, France}
\altaffiltext{8}{European Southern Observatory, Germany}
\altaffiltext{9}{Massachusetts Institute of Technology, Cambridge, MA}
\altaffiltext{10}{IMEP-INPG, Grenoble, France}


\begin{abstract}
  We report the first spatially-resolved observations of the
  spectroscopic binaries $\lambda$~Vir and WR~140, which includes the
  debut of aperture-synthesis imaging with the upgraded
  three-telescope IOTA interferometer.  Using IONIC-3, a new integrated optics
  beam combiner capable of precise closure phase measurement, short
  observations were sufficient to extract the angular separation and orientation of
  each binary system and the component brightness ratio.  Most
  notably, the underlying binary in the prototypical colliding-wind
  source WR~140 (WC7 + O4/5) was found to have a separation of
  $\sim$13~milli-arcseconds with a position angle consistent with the
  images of the 2001 dust shell ejection only if the Wolf-Rayet star
  is fainter than the O star at 1.65$\mu$m.  We also highlight
  $\lambda$~Vir whose peculiar stellar properties of the Am star
  components will permit direct testing of current theories of tidal
  evolution when the full orbit is determined.
\end{abstract}




\section{Introduction}
Only two optical (visible and infrared) interferometers have published
results combining three or more separated telescopes, the Cambridge
Optical Aperture Synthesis Telescope \citep{baldwin1996} and the Navy
Prototype Optical Interferometer \citep{benson1997}.  Combining more
than two telescopes is crucial in order to measure the {\em closure
phase}, a phase quantity that is uncorrupted by atmospheric
turbulence and necessary for image reconstruction in ``very long
baseline interferometry'' \citep{jennison58, monnier_mss}; indeed,
closure phase lies at the heart of so-called ``self-calibration''
techniques widely used in the radio \citep[e.g.,
][]{readhead1980,cw81}.

Here we present the first scientific results of the upgraded
3-telescope IOTA interferometer \citep{traub2003}, with the new
integrated-optics combiner IONIC-3 \citep{berger2003} for precise
measurements of visibilities and closure phases.  IOTA has much
greater near-infrared sensitivity (demonstrated H mag 7) than the
other imaging arrays, allowing new observational programs including
imaging of young stellar objects, mass determinations of low-mass
binaries, and mapping photospheric surface structures of red giants.
After describing our novel instrumentation, we present the first
resolved observations for the A-star binary $\lambda$~Vir and
colliding wind system WR~140 and discuss the significance of our
findings.  Here we alert the astronomical community to the need for
radial velocity follow-up of these specific sources as well as to the
new research opportunities available with the IOTA3 array.

\section{Observations and Analysis}
\label{section:observations}
\label{analysis}
All the data presented were obtained using the IOTA (Infrared-Optical
Telescope Array) interferometer \citep{traub2003}, which is operated
by a consortium of institutions, most notably the Smithsonian
Astrophysical Observatory and the University of Massachusetts at
Amherst.  The 0.45-m telescopes are movable among 17 stations along 2
orthogonal linear arms (telescopes A \& C can move along the 35-m
northeastern arm, while Telescope B moves along the 15-m southeastern
arm), allowing an aperture 35m$\times$15m to be synthesized
(corresponding to an angular resolution of
$\sim$5$\times$12~milliarcseconds at 1.65$\mu$m).  Capping a long
development \citep[e.g.,][]{schloerb1990}, the third telescope (``C'')
was fully incorporated on UT~2002~February~23 and now allows
visibilities on three baselines and one closure phase to be measured
simultaneously. The initial observations used a prototype integrated
optical (IO) combiner \cite[e.g.,][]{kern1997, berger2001} which
operated only with polarized light.  The upgraded combiner IONIC-3
corrected this deficiency in 2002 November, and all observations
presented here were obtained in unpolarized light.

We only just introduce IONIC-3 \citep{berger2003} here;
an engineering paper with 
detailed description of the IO component and its performance is
in preparation (Berger et al. 2004).  Light from each telescope is
focused into a dedicated single-mode fiber, and the three fibers are
aligned using a silicon v-groove array mated to the planar waveguides
on the IO device.  The optical circuit acts to split the light from
each telescope before recombining each telescope pair (AB,BC,AC)
at 3 IO couplers.  This ``pair-wise'' combination scheme leads to
6-interferometric channels (two for each baseline), where the
interference fringes are temporally-modulated by scanning piezo
mirrors placed in two of the telescope beams; a sensitive
HgCdTe (Rockwell PICNIC) array (Pedretti et al. 2004a, submitted)
detects the signals, and data are recorded by a vme-based
control system linked to a Sun workstation.

\begin{figure}[hbt]
\begin{center}
\includegraphics[angle=90,clip]{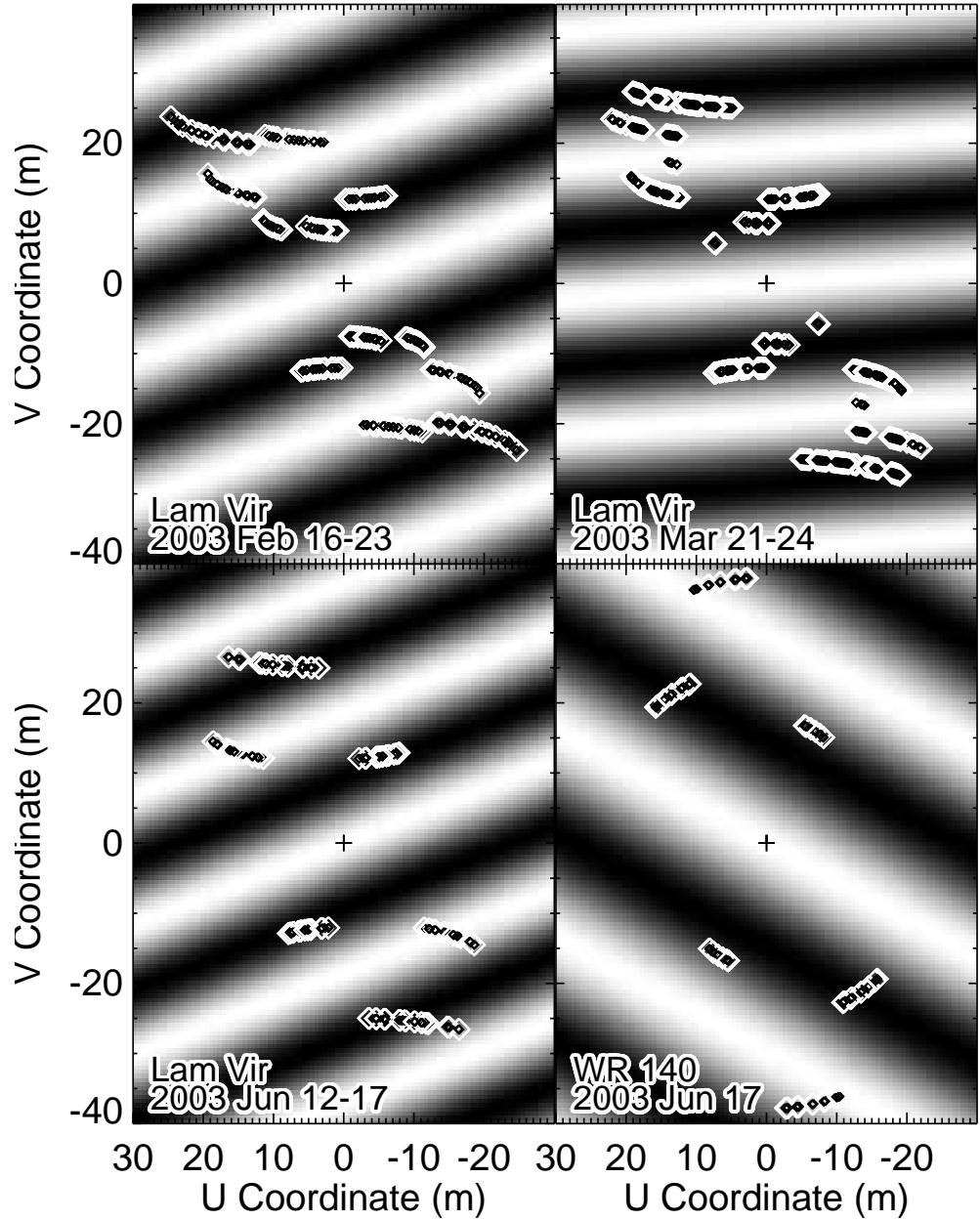}
\caption{(u,v)-plane coverage of science binaries.  Each 
diamond symbol represents a single observation with the IOTA3
interferometer, corresponding to approximately 100-200 interferogram
scans. The overlaid image in each panel represents the \vis~of the best fit binary
model for each epoch discussed in text.
\label{fig_uvcov}}
\end{center}
\end{figure}

We followed established observing procedures, as outlined in previous
IOTA publications \citep{foresto1997, millangabet2001}: observations
of target and nearby calibrator stars were interspersed to calibrate
slowly varying system visibilities and instrumental closure phases.
The science targets presented here were observed
using a standard H band filter ($\lambda_0 = 1.65\mu$m, $\Delta\lambda
= 0.30\mu$m); Table~\ref{table_iota} contains other observing details.
Figure~\ref{fig_uvcov} shows the
(u,v)-coverage obtained during each epoch for each science target.

Reduction of the \vis~data was carried out using an {\em
Interactive Data Language} (IDL) implementation of the method outlined
by \citet{foresto1997}.  In short, we measured the power spectrum of
each interferogram (proportional to the target \vis), after correcting
for intensity fluctuations and subtracting out bias terms from read
noise, residual intensity fluctuations, and photon noise
\cite[e.g.,][]{perrin2003a}.  To assure good closure phase
measurement, we required that interferograms be detected on $\ge2$
baselines, a condition nearly always maintained by a realtime fringe
packet ``tracker'' (Pedretti et al. 2004b, in preparation).  Lastly,
the data pipeline applied a correction for the variable flux ratios
for each baseline by using a transfer matrix \citep[based on
single-beam flux measurements; e.g.,][]{foresto1997,monnier2001}.  We
have studied the absolute calibration accuracy by observing single
stars of known size and found a systematic error of
$\Delta$\vis$=0.10$ (e.g., 5\% error in visibility for unresolved
sources) under the worst observing conditions; this error has been
combined (in quadrature) with the statistical error for fitting
purposes.  Improved analysis methods should achieve $\simle$1\% errors
eventually, since we are using single-mode fibers.

We followed the method of \citet{baldwin1996} for calculating the
complex triple amplitude in deriving the closure phase, explicitly
guaranteeing ``fringe frequency closure'' ($\nu_{AB} + \nu_{BC} +
\nu_{CA} = 0$).  While photon-noise bias correction is not necessary
for the IONIC-3 (``pair-wise'') combiner, the instrumental closure
phase must be determined empirically.  The miniature dimensions of the
IO component have minimized drifts to less than 1~degree over many
hours; chromaticity effects, however, limit our absolute precision
when the calibrator and source are not of the same spectral type due
to different ``effective wavelengths.''  Engineering tests indicate
that the instrumental closure phase ($\Phi_{CP}$) varies
systematically by $1.4\pm0.3$~degrees between a hot star (B8) and a
cool star (M3) when using the broadband H filter.  While we minimize
these errors by using calibrators matched to the spectral type of the
target, a conservative systematic error of
$\Delta\Phi_{CP}=0.75$~degs was adopted here to approximate our
expected residual level of miscalibration.  Lastly, the sign of our
closure phase ($\Phi_{AB}+\Phi_{BC}+\Phi_{CA}$ for baseline triangle
connecting telescopes A$\rightarrow$B$\rightarrow$C$\rightarrow$A) was
calibrated using the well-known binary stars Capella and Matar
\citep{hummel1994,hummel1998}.

\section{Results and Discussion}
\label{section:results}
With calibrated \vis~and closure phases for each target, a static
binary star model (separation, position angle, brightness ratio) was
fitted to the data for each given epoch, assuming the calibration errors
given in the previous section and ignoring bandwidth-smearing effects.
Table~\ref{table_params} contains the results of our fits based on a
maximum-likelihood approach giving equal weights to the visibility
data and closure phases.
Explicitly, the probability of a given set of parameters was
approximated by: ${\rm Probability} \propto exp( - \frac{\chi^2_{{\rm
Vis}^2}} {{\rm minimum}(\chi^2_{{\rm Vis}^2})} - \frac{\chi^2_{{\rm
CP}}} {{\rm minimum}(\chi^2_{{\rm CP}})})$. This normalization is
equivalent to increasing the data errors to force the
$\chi^2$/DOF$=1$, and conservatively estimates
errors in the context of an incomplete error model.
The most probable values with 1-sigma
errors bars (68.4\% confidence interval) were calculated for each
parameter (separation, position angle, brightness ratio) by
integrating over the other variables, a standard practice referred to
as {\em marginalizing} over these variables.

\begin{figure}[hbt]
\begin{center}
  \includegraphics[angle=90,clip,width=4.0in]{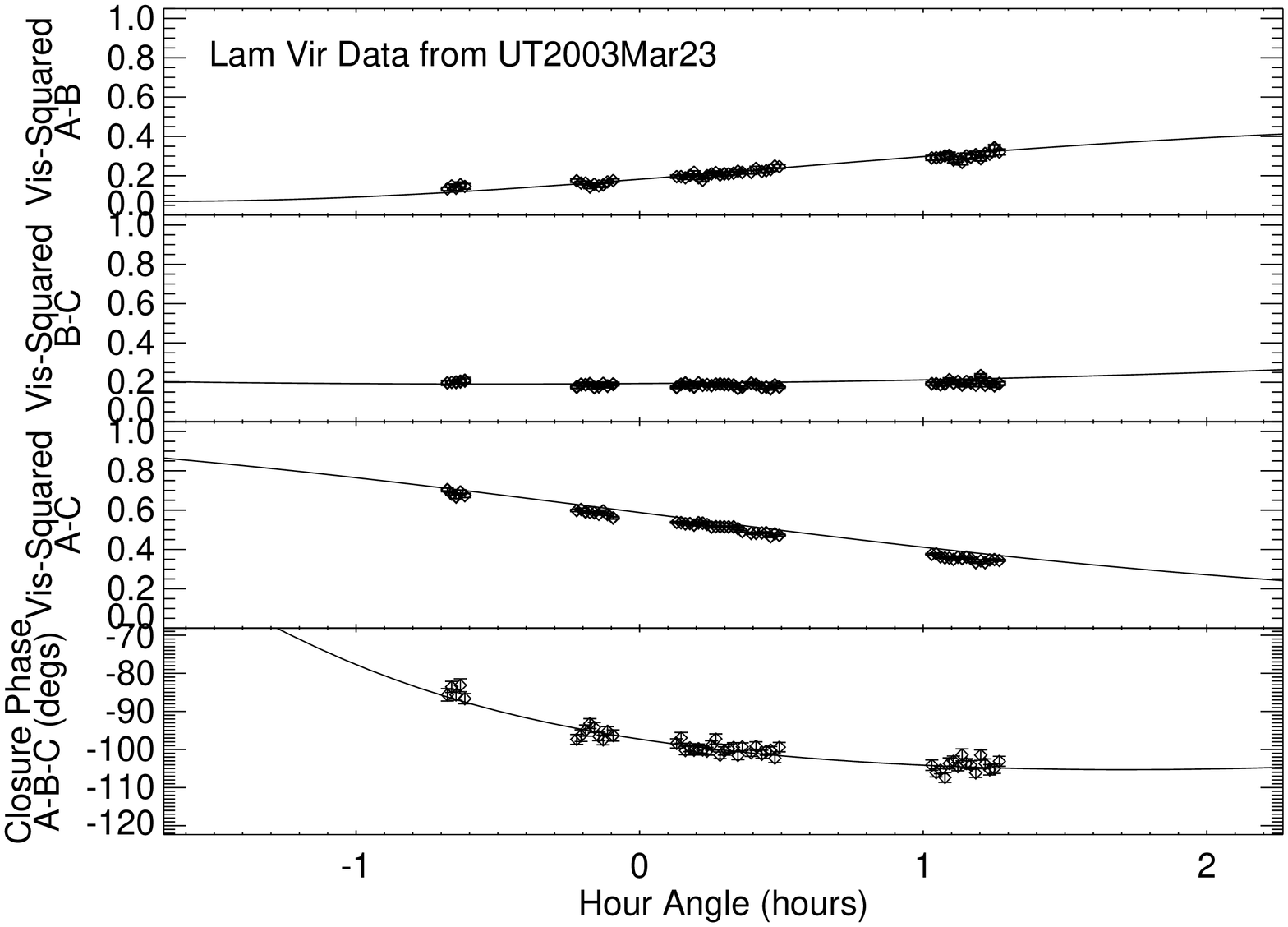}
\vspace{.25in}
  \includegraphics[angle=90,clip,width=4.0in]{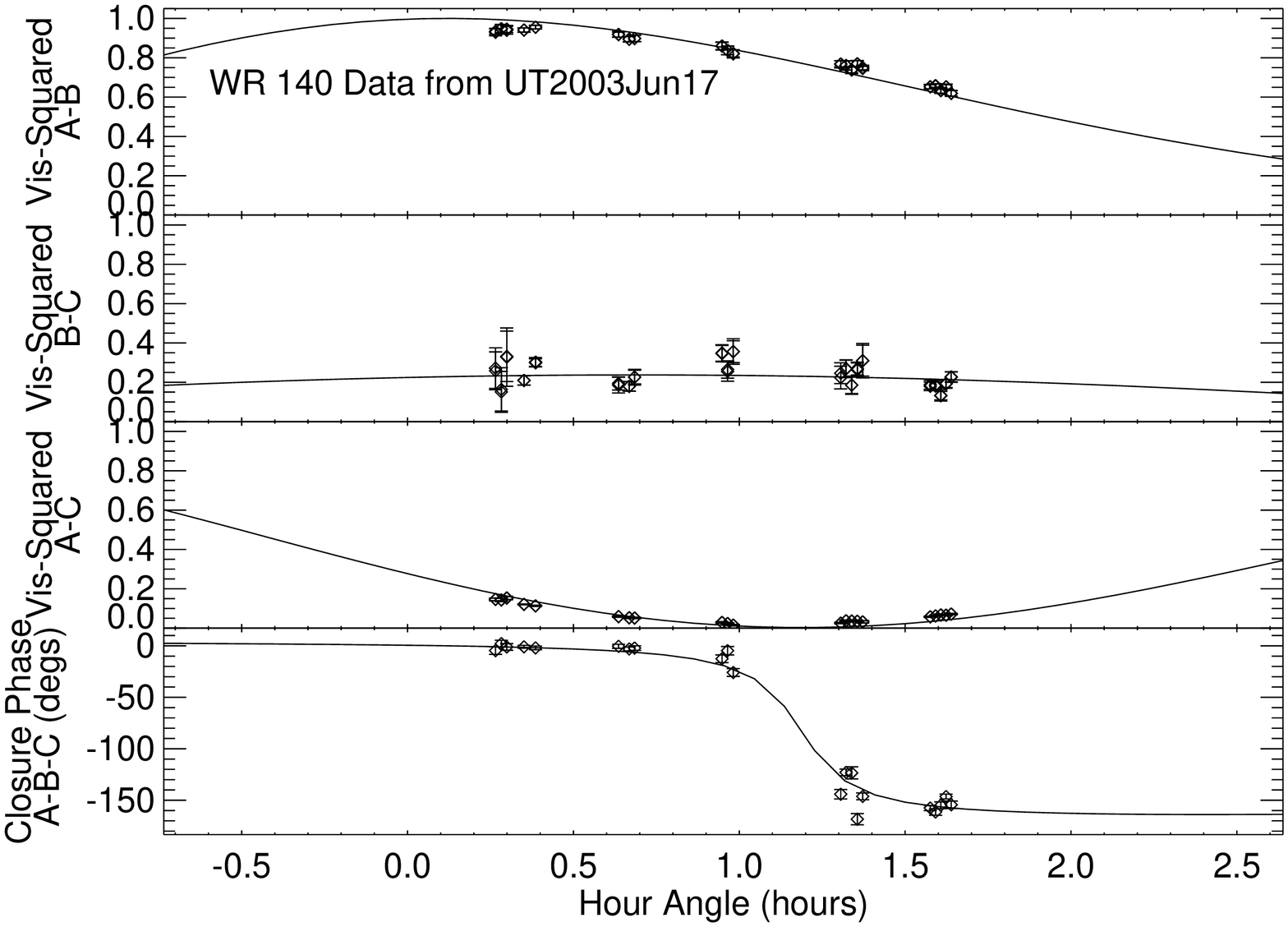}
  \caption{{\em (top panel):} This figure shows example binary
    model fits for $\lambda$~Vir on UT~2003~March~23. The top three panels
    show the \vis~for baselines connecting telescopes A-B, B-C, and
    A-C, respectively.  The bottom panel shows the calibrated closure
    phase for the A-B-C triangle.  {\em (bottom panel):} Same as top
    panel for UT~2003~June~17 observations of WR~140. These curves
    represent the best-fit model parameters as described in \S\ref{section:results}.
For all panels, the
    error bars shown represent only the statistical error and do not
    include the systematic error discussed in 
    \S\ref{analysis}. 
\label{fig_fits}}
\end{center}
\end{figure}

Based on their brightness and color temperatures, the component stars
of both binaries are expected to be unresolved here ($\simle 0.5$~mas);
this assumption, as well as neglecting bandwidth-smearing, may not
remain valid if the data precision is improved significantly.  The
\vis~and closure phases for a subset of the data are shown in
Figure~\ref{fig_fits} along with the best-fitting model prediction. 
We note that $\lambda$~Vir, which has a $\sim207$~day period, is not
completely static during each observing ``epoch.''
With complete orbital phase coverage,  
this deficiency will be corrected by
solving for the
full {\em orbital elements} by
fitting directly to the visibilities, closure phases, and radial velocities,
a procedure outlined in detail by \citet{hummel1998} and \citet{boden1999}.

\begin{figure}[hbt]
\begin{center}
\includegraphics[angle=0,clip,width=5in]{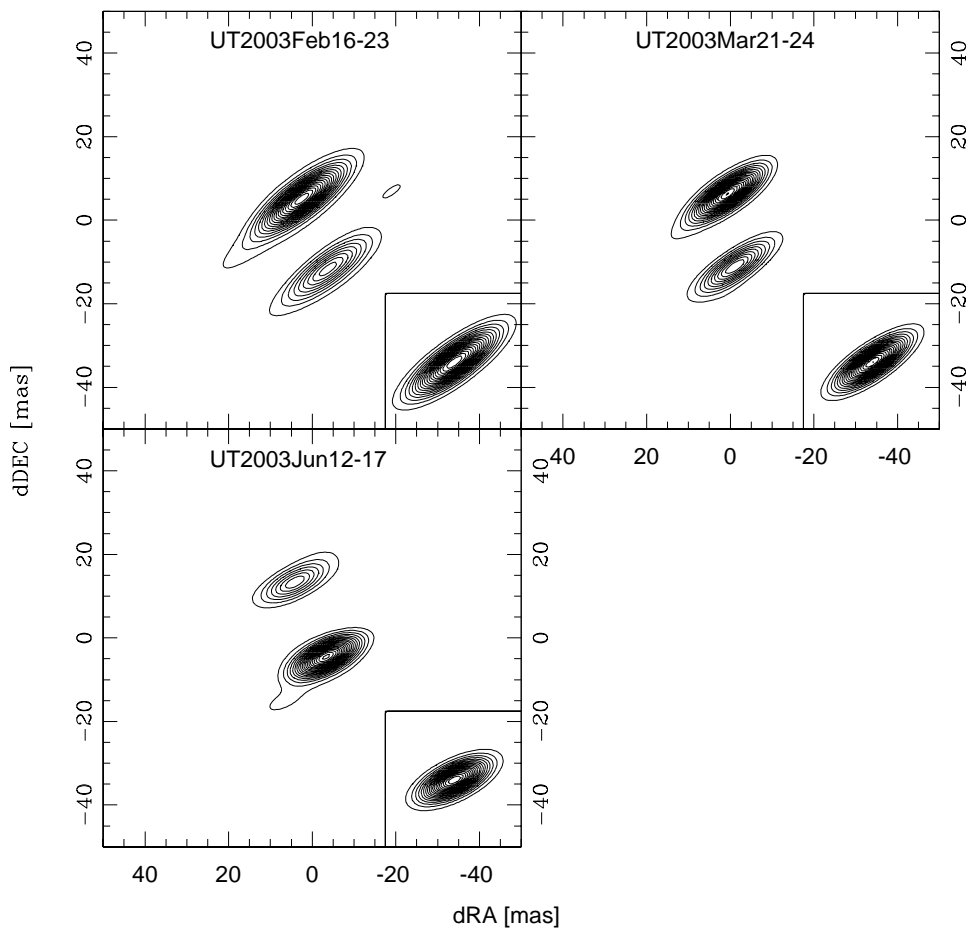}
\caption{First aperture synthesis images from the upgraded
IOTA 3-telescope interferometer, depicting
$\lambda$~Vir based on 1.65$\mu$m observations at three separate 
epochs. The binary components are well-resolved from each other but individually are
unresolved; orbital motion is clearly evident, showing nearly a $\sim$180\arcdeg~ 
rotation between the first and last epoch.  Each panel includes
a contour map showing
5\% intervals in surface brightness scaled to the peak,
as well as the 
the CLEAN restoring beam as inset.
North is up and East is left.
\label{fig_image}}
\end{center}
\end{figure}

The radio community developed interferometric imaging techniques 
(e.g., hybrid mapping) capable of reobtaining lost individual
phases \citep[e.g.,][]{haniff1987}. As demonstrated by
\citet{baldwin1996}, these methods can be applied successfully
to optical interferometers as well. We present the first aperture
synthesis images from IOTA3 in Figure~\ref{fig_image}, depicting
$\lambda$~Vir for each epoch. The images are constructed iteratively
starting with zero initial phases and using the measured closure phases
for self-calibration. CLEAN deconvolution was applied
using an elliptical Gaussian CLEAN beam.  Full details of the
image-making process for this source (as well as the binary
Capella) can be found in \citet{kraus2003}. The presented maps are
consistent with the binary parameters
found through direct fitting, and demonstrate the imaging capability
of the IOTA3+IONIC-3 system, currently the most sensitive infrared
imaging interferometer in the world.

With this new imaging capability, we begin to study binary
stars which have never been resolved before and which represent unique stellar systems whose
properties can shed new light on rare and/or short-lived phases of
stellar evolution.


$\lambda$~Vir is a well-known spectroscopic binary \citep[period
207~days,][]{stickland1976} with similar components of spectral type
early Am \citep[both are metallic-lined,][]{abt1961}.  The stars
exhibit the curious property that one is sharp-lined and the other
broad-lined, which makes them easily distinguishable
spectroscopically. IOTA3 observations will
allow a determination of the absolute properties of the components
once the full orbit is established.  Fundamental stellar parameters,
such as the system distance (independent of Hipparcos) and
the individual masses and
absolute near-IR luminosities, 
can be determined through a combination of the radial velocity orbit,
astrometric orbit, and brightness ratio.
Further, these properties can be
compared directly against models of stellar evolution to derive the
age of the system.  The very different projected rotational
velocities of the components is of special interest for
these Am stars, enabling unique tests of 
current theories of tidal evolution \citep[][and references
therein]{tassoul1997,zahn1989}.

WR~140 is a prototypical colliding-wind binary source consisting of a
Wolf-Rayet (WC7) and an O4-5 star in a highly-eccentric $\sim$7.9~yr orbit,
and has been extensively studied in the radio, infrared, optical,
ultra-violet, and X-rays \citep[e.g.,][]{moffat87,williams90,wb95}.
Notably, infrared emission appears and fades in conjunction with
periastron passage, explained by transient dust formation in the dense
shock-compressed gas at the wind-wind interface
\citep{williams90,usov91,tuthill1999b}.  This process was captured 
in
a series of high-resolution near-infrared images by
\citet{monnier2002b}, revealing an eastward moving arc of dust
consistent with the O-star being roughly east of the WR star at
periastron.  The O-star is
expected to be on (nearly) the opposite side of the WR star now as
compared to periastron, which can be reconciled with the position angle
found here (PA 152\arcdeg) only if the WR star is fainter than
the O-star at 1.65$\mu$m \citep[as expected by some workers; 
e.g.][]{williams90}.  We intend to combine future high-precision
interferometer observations with new radial velocity data
\citep[e.g.,][]{marchenko2003} to derive component masses and a
distance estimate, fundamental parameters for these systems needed for
definitive interpretation of data across the electromagnetic spectrum.

\section{Conclusions}
\label{sectionc:conclusions}
We have reported first scientific results from the IOTA3
interferometer: newly-resolved separation, position angle, and
flux ratio information for double-lined spectroscopic binary stars
$\lambda$~Vir and WR~140. Using a preliminary data reduction pipeline,
calibration is sufficient for model-fitting and aperture-synthesis
imaging using hybrid mapping techniques.  New observations are being
pursued in order to derive precise component masses and distance
estimates by combining interferometric and radial velocity data.  All
the visibility and closure phase data here have been converted to the
new FITS format for Optical Interferometry data (OI-FITS)
\footnote{http://www.mrao.cam.ac.uk/$\sim$jsy1001/exchange/ } and are
available upon request.

\acknowledgments {The authors gratefully acknowledge critical support
  from SAO, NASA, and NSF (AST-0138303).  EP was supported by a SAO
  Predoctoral fellowship, JDM by a CfA fellowship, and RM-G, J-PB, and
  SR through NASA Michelson Fellowships. GT acknowledges partial
  support from NASA's MASSIF SIM Key project (JPL 1240033). In
  addition, we acknowledge useful contributions from
  B. Arezki, A.  Delboulbe, C. Gil, S.  Gluck, E.  Laurent,
  R. B. Metcalf, and E. Tatulli.  IONIC-3 was
  developed by LAOG and LETI in the context of the IONIC collaboration
  (LAOG, IMEP, LETI), funded by the CNRS and CNES (France). This
  publication makes use of data products from the 2MASS survey, which
  is a joint project of the University of Massachusetts and
  IPAC/Caltech, funded by NASA and the NSF.  This research has also
  used the SIMBAD database, NASA's ADS Service, and services of the
  Michelson Science Center at Caltech (http://msc.caltech.edu).  }

\bibliographystyle{apj}
\bibliography{apj-jour,IOTA3,Review,Review2,Thesis,WR140}


\clearpage

\begin{deluxetable}{lll}
\tablecaption{Observing Log for $\lambda$~Vir and
WR~140
\label{table_iota}}
\tablehead{
  \colhead{Date}     & \colhead{Interferometer}  & \colhead{Calibrator Names}\\
  \colhead{(UT)} & \colhead{Configuration\tablenotemark{(a)}} & \colhead{(Adopted UD
    Diameter)} } \startdata
\multicolumn{3}{c}{$\lambda~Vir$ (A1V + A);  H mag $=4.28\pm$0.21 } \\
2003 Feb 16,17  & A35-B05-C10 & HD~126035 (G7III, 0.78$\pm$0.24~mas) \\
&&HIP~71957 (F2III, 1.20$\pm$0.22~mas) \\
2003 Feb 20,22,23 & A25-B15-C10 &  HD~126035 \\
2003 Mar 21 & A35-B07-C25 &  HD~126035 \\
2003 Mar 22 & A35-B07-C10 & HD~126035; HD~158352 (A8V, 0.44$\pm$0.10) \\
2003 Mar 23-24 & A35-B15-C10 &  HD~126035; HD~158352 \\
2003 Jun 12-17 & A35-B15-C10 & HD~126035 \\
\hline
\multicolumn{3}{c}{WR~140 (WC7 + O4/5);  H mag $=5.429\pm$0.023 } \\
2003 Jun 17 & A35-B15-C10 & HD~192985 (F5IV, 0.46$\pm$0.10 mas)\\
&& HD~193631 (K0, 0.44$\pm$0.10 mas)
\enddata
\tablenotetext{~}{Notes: H-band photometry from 2MASS All-Sky Release; calibrator
diameters estimated using {\em getCal} software (http://msc.caltech.edu).}
\tablenotetext{a}{Configuration refers to the location of telescopes A,B,C on
the NE, SE and NE arm respectively; see \S\ref{section:observations} for more details.}

\end{deluxetable}
\clearpage
\begin{deluxetable}{llccccc}
  \tablecaption{Binary Parameters for
    $\lambda$~Vir and WR~140 at 1.65$\mu$m
\label{table_params}}
\tablehead{
  \colhead{Source}     & \colhead{Epoch}  &  \colhead{Separation} &
  \colhead{Position Angle\tablenotemark{(a)}} & \colhead{Flux} & \multicolumn{2}{c}{$\chi^2$/DOF\tablenotemark{(b)}} \\
   & \colhead{(U.T.)} & \colhead{(milliarcseconds)} & \colhead{(degrees)} & 
  \colhead{Ratio} &\colhead{$V^2$} & \colhead{CP}} 
\startdata
$\lambda$~Vir & 2003 Feb 16-23 & $18.2\pm 0.6$ & $200\pm 3$  & $1.69^{+0.29}_{-0.21}$  & 0.45 & 12.8  \\
 & 2003 Mar 21-24 & $17.8\pm 0.5$ & $184\pm 3$ & $ 1.72\pm 0.07$ & 0.33 & 2.2\\ 
 & 2003 Jun 12-17 & $19.5\pm 0.6$ & $22\pm 3$ & $2.05^{+0.50}_{-0.35}$& 0.31& 4.8\\
WR~140 & 2003 Jun 17 & $12.9^{+0.5}_{-0.4}$ & $151.7^{+1.8}_{-1.3}$ & $1.10^{+0.06}_{-0.05}$ & 0.13 & 4.7 \\
\enddata
\tablenotetext{a}{Position angles
(East of North)
are measured from the bright to the faint component.}
\tablenotetext{b}{The $\chi^2$ per DOF (degree of freedom), the {\em reduced} $\chi^2$, is reported for the model fit to the
\vis~and closure phase (CP) measurements separately.}
\end{deluxetable}
\clearpage

\clearpage

\clearpage

\end{document}